\documentclass[ %
 reprint,
superscriptaddress,
 amsmath,amssymb,
 aps,
prb,
]{revtex4-1}

\usepackage{graphicx}
\usepackage{dcolumn}
\usepackage{bm}
\usepackage{subfigure}
\usepackage{epstopdf}
\usepackage{lettrine}

\begin{document}

\preprint{APS/123-QED}

\title{Demonstration of electron focusing using electronic lenses in low-dimensional system}

\author{Chengyu Yan}
\email{uceeya3@ucl.ac.uk}
\affiliation{%
	London Centre for Nanotechnology, 17-19 Gordon Street, London WC1H 0AH, United Kingdom\\
}%
\affiliation{
	Department of Electronic and Electrical Engineering, University College London, Torrington Place, London WC1E 7JE, United Kingdom
}%
\affiliation{%
	Micronova, Aalto University, Tietotie 3, Otaniemi, Espoo, 02150,Finland\\
}%
\author{Michael Pepper}
\affiliation{%
	London Centre for Nanotechnology, 17-19 Gordon Street, London WC1H 0AH, United Kingdom\\
}%
\affiliation{
	Department of Electronic and Electrical Engineering, University College London, Torrington Place, London WC1E 7JE, United Kingdom
}%
\author{Patrick See}
\affiliation{%
	National Physical Laboratory, Hampton Road, Teddington, Middlesex TW11 0LW, United Kingdom\\
}%
\author{Ian Farrer}
\affiliation{%
	Department of Electronic and Electrical Engineering, University of Sheffield, Sheffield, S1 3JD, United Kingdom\\
}%
\author{David Ritchie}
\affiliation{%
	Cavendish Laboratory, J.J. Thomson Avenue, Cambridge CB3 OHE, United Kingdom\\
}%
\author{Jonathan Griffiths}
\affiliation{%
	Cavendish Laboratory, J.J. Thomson Avenue, Cambridge CB3 OHE, United Kingdom\\
}%

\date{\today}
             
\begin{abstract}

\section*{Abstract}
	
We report an all-electric integrable electron focusing lens in n-type GaAs. It is shown that a pronounced focusing peak takes place when the focal point aligns with an on-chip detector. The intensity and full width half maximum (FWHM) of the focusing peak are associated with the collimation of injected electrons. To demonstrate the reported focusing lens can be a useful tool, we investigate characteristic of an asymmetrically gate biased quantum point contact with the assistance of focusing lens. A correlation between the occurrence of conductance anomaly in low conductance regime and increase in FWHM of focusing peak is observed. The correlation is likely due to the electron-electron interaction. The reported electron focusing lens is essential for a more advanced electron optics device.     
 
\end{abstract}

\maketitle

\section*{Introduction}

During the past several years, electronic and optical technologies have seen many encouraging developments. On the electronic end, emergent devices such as spintronics\cite{CFX05,WPI10,BDS12} and valleytronics\cite{RTB07,SYC16} have been proposed and realized;  similarly, devices such as coherent optical memory\cite{HDD04} and optical qubit\cite{PAL04} have laid the foundation of optical quantum computation. Integrating optical and electronic properties into a single platform or electron optics will provide a unique system for investigating many phenomena emerging from their fusion.  The wave nature of electrons in low dimensions, especially ballistic electrons in clean semiconductors, could be exploited for geometrical optical phenomena.  In order to realize the potential of electron optics, it is necessary to establish a mapping between the fundamental optical components and their electronic counterparts. In this regard, it is well known that quantum point contacts (QPCs) or other low-dimensional electron sources are equivalent to a coherent optical source\cite{TLS00}; electronic spin polariser is inspired by an optical polariser\cite{DD90}; electronic cavity/mirror shares functional similarities with the optical ones\cite{DTW01}.  Electron focusing with the electrostatic lens has been demonstrated with a  double-concave lens\cite{SHU90,SSB90}. However, the primary parameter that determines the focusing profile has not been addressed.

In the present work, we demonstrate an all-electric electron focusing lens with a more intuitive design and identify the main factor that shapes the focusing profile. A focusing peak occurs whenever the focal point, which is gate-voltage tunable, spatially aligns with the on-chip detector. It is found that the focusing profile is closely associated with the collimation of injected electrons. To give an example of the potential usage of the electronic focusing lens, we utilized this technique to investigate the characteristic of an asymmetrically gate biased QPC. A correlation between the occurrence of conductance anomaly in low conductance regime and increase in FWHM of focusing peak is observed. The correlation is likely due to electron-electron interaction.

\begin{figure}
	
	\includegraphics[height=3.2in,width=3.2in]{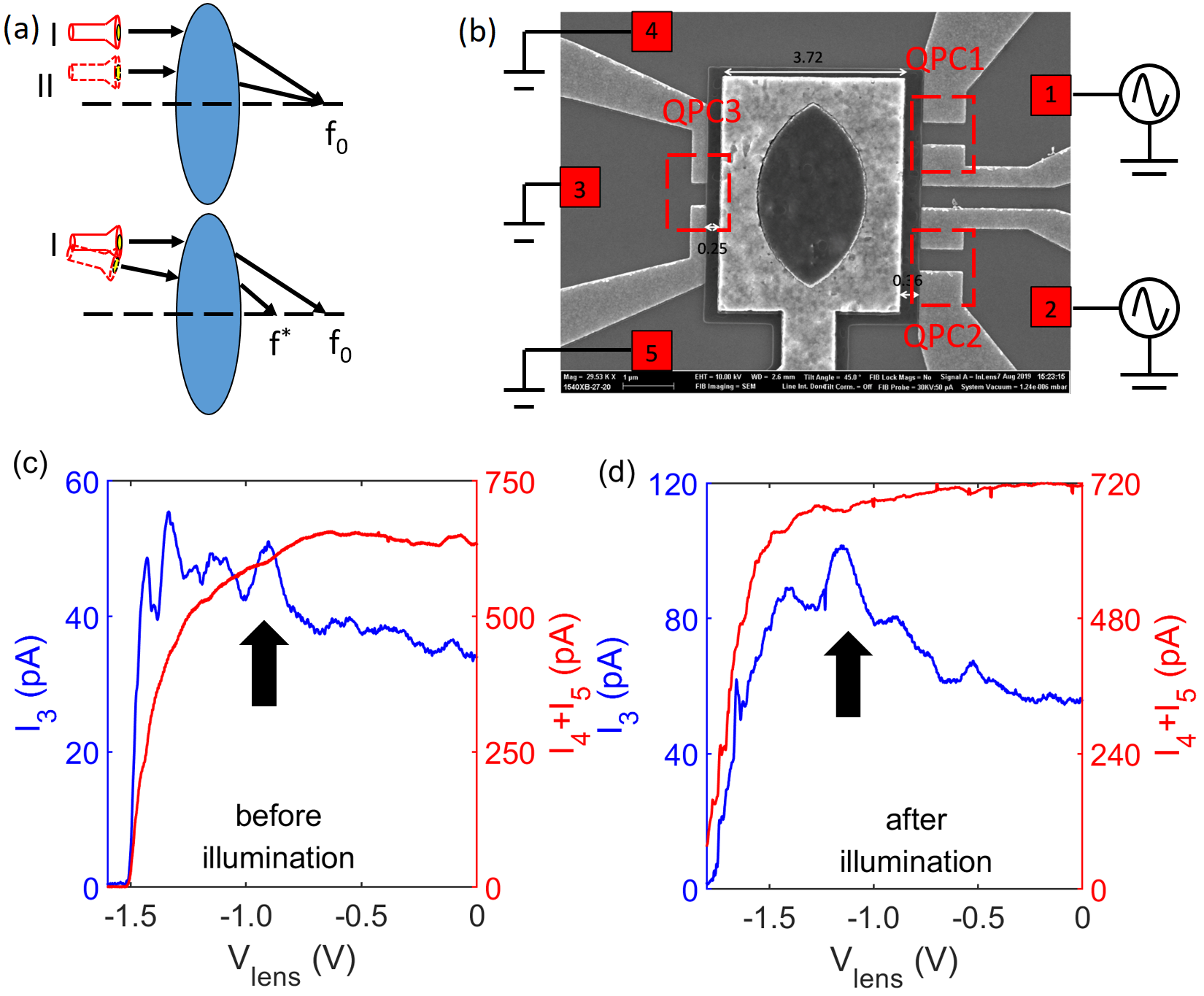}
	
	\caption{Schematic of experiment setup and representative results of electronic focusing. \textbf{(a)} Typical functionality of an optical double-convex lens. If the incident light is parallel with the primary axis of the lens, then it would be guided to the focal point $f_0$ (upper panel); on the other hand, light with a non-zero incident angle with respect to the primary axis would be guided to $f^\ast$ instead of $f_0$. \textbf{(b)} SEM image of the experiment setup. The dimension of the split gates and lens gate of the imaged device are the same as those used in the experiment; the difference relies on that the gap between the split gates and lens gate are 300 nm for the imaged device, and 100 nm (dev A) and 150 nm (dev B) for the measured devices. QPC1 and 2 are injectors; QPC3 is used as detector. The top-gate with a hollow area is referred as lens gate, and it is patterned over a PMMA layer. Squares 1-5 at the edges of the mesa represent Ohmic contacts. An enlarged image, Fig. S1, can be found in the supplementary information. \textbf{(c)} and \textbf{(d)} show representative results before and after the sample was illuminated with a red LED. QPC 1-3 were set to G$_0$. $I_3$ represents the signal through detector where a noticeable enhancement, marked by the bold black arrow, was observed; $I_4$+$I_5$ is the signal drained to Ohmic contact 4 and 5. It is important to point out that, $I_3$ and $I_4$+$I_5$ were measured simultaneously. It should be commented that the focusing peak is observable either use a single injector or both.  }
	
	\label{fig:1}
	
\end{figure} 

\begin{figure}
	
	\includegraphics[height=2.0in,width=3.4in]{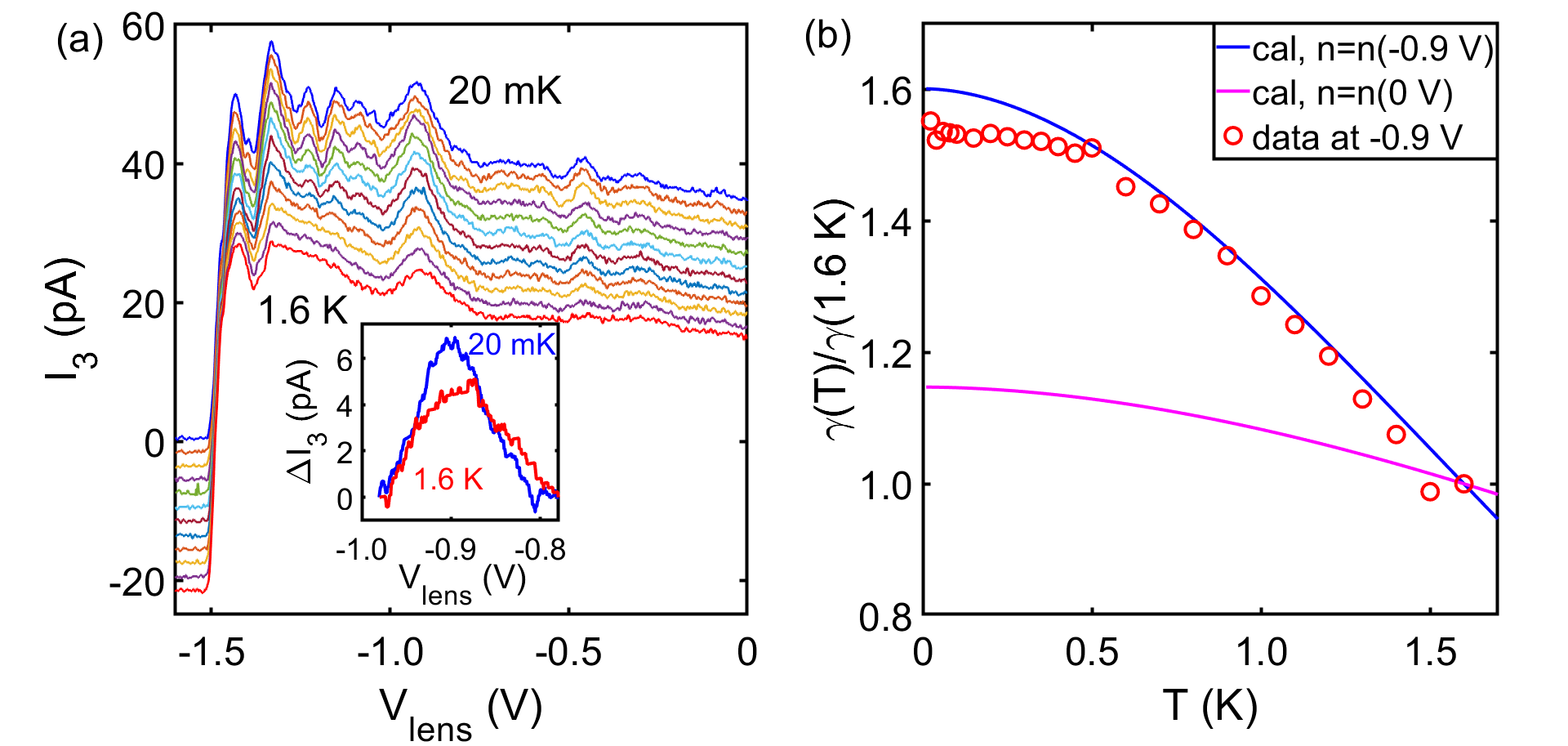}
	
	\caption{Temperature dependence of focusing signal. \textbf{(a)} QPC 1-3 were set to G$_0$, the lattice temperature was incremented from 20 mK to 1.6 K. Data have been offset vertically for clarity. Inset shows a zoom-in of focusing peak at 20 mK and 1.6 K after removing a linear background. \textbf{(b)} Normalized peak intensity $\gamma(T) / \gamma (1.6K)$ as a function of temperature. $\gamma(T)$ was determined by subtracting a linear background within the vicinity of focusing peak. The solid blue and magenta lines show the calculated upper and lower bound of peak intensity using Eq.2 and 3 without adjustable parameters. Details about the calculation can be found in the main text.  }
	
	\label{fig:2}
	
\end{figure}

\begin{figure}
	
	\includegraphics[height=2.4in,width=3.2in]{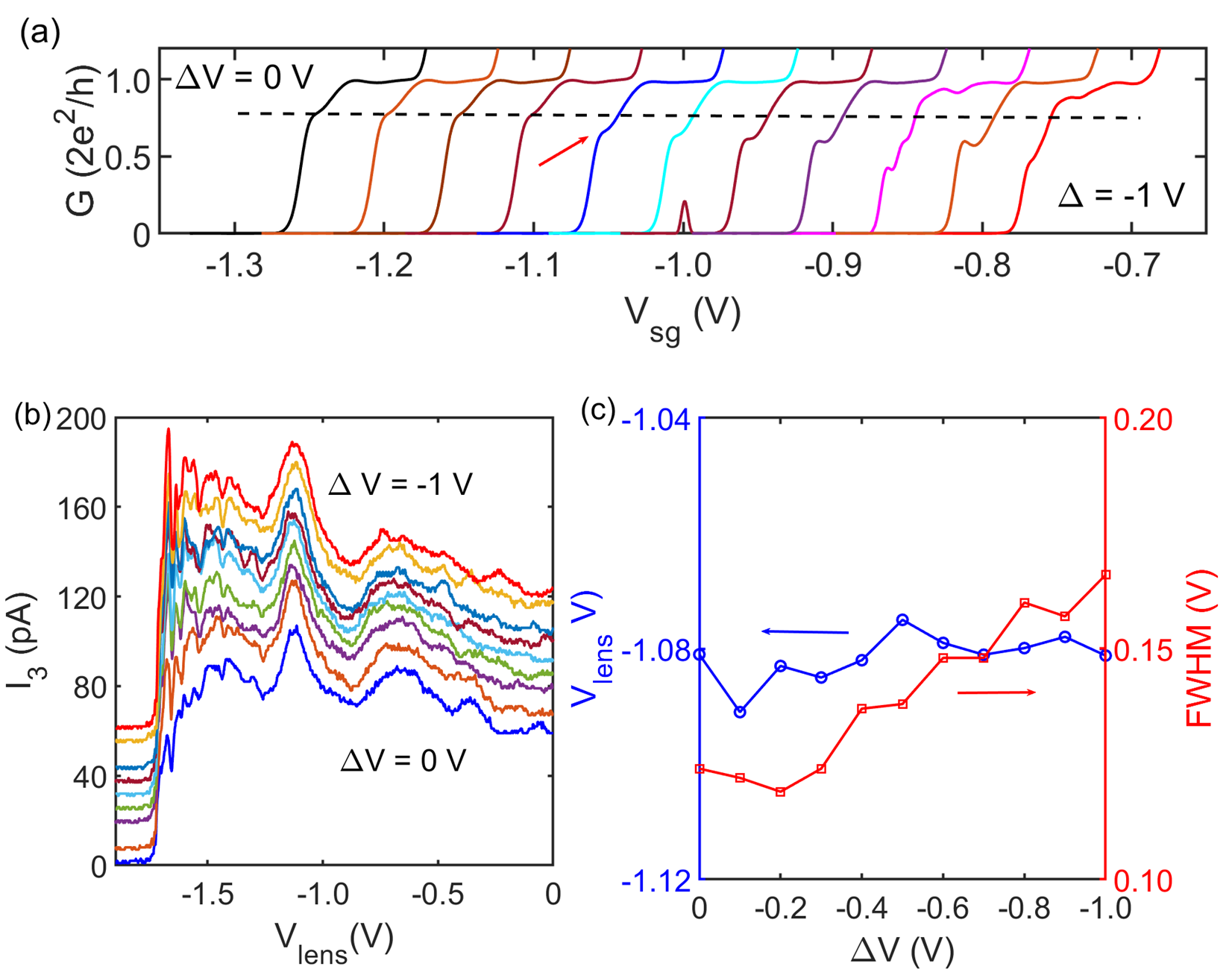}
	
	\caption{Focusing peak as a function of asymmetric gate bias after illumination. \textbf{(a)} Conductance characteristic of QPC1 with different asymmetric gate bias $\Delta V$; gate voltage $V_{sg}$ was applied to the bottom arm of the split gate [refers to Fig.1 (b)], $V_{sg}+\Delta V$ was applied to the upper arm. Series resistance has not been removed. The red arrow highlights the occurrence of a sub-0.7-anomaly. \textbf{(b)} Focusing peak with different $\Delta V$ applied to QPC1. QPC1 was set to G$_0$,  the gate voltage applied to the two arms of QPC1 were calibrated according to \textbf{(a)}, QPC2 was not used; QPC3 was under symmetric gate bias and fixed at G$_0$. Data in \textbf{(b)} have been offset vertically for clarity. \textbf{(c)} Peak position (blue) and FWHM (red) as a function of $\Delta V$. It is necessary to mention a background has been removed in determining the FWHM. The background can be determined from two ways: 1. make a polynomial fitting in the vicinity of the focusing peak, as shown in supplemental Fig. 1; 2. scales the data with both Ohmic 4 and 5 floating which does not include the correction due to focusing process, so that at zero lens gate voltage (the scaled) $I_3$ is the same when Ohmic 4 and 5 are floating and grounded. The two methods lead to a similar conclusion.   }
	
	\label{fig:3}
	
\end{figure}       

\section*{Experiment}

To realize an electronic analogue of an optical focusing scheme, it requires both collimated electron source and electron focusing lens [Fig. 1(a)]. The highly collimated ballistic electrons\cite{MSB90,STO92} are injected by two QPCs, i.e., QPC1 and 2 as shown in Fig. 1(b), whereas QPC3 functions as a detector. Electron injection angle concentrates at 0$^\circ$ when the QPC is confined to low conductance regime, but always has a finite angular spread\cite{MSB90,STO92}. Before being collected at QPC3, collimated electrons pass through an electronic focusing lens defined via a top gate encapsulating a double-concave shaped hollow regime [[Fig. 1(b)]; see supplementary information note 1 and 2 for comments on lens design\cite{SI19}]. Electron refraction follows the Snell's law\cite{SHU90,SSB90}. The relative refractive index $N_r = \sqrt{\frac{n_1}{n_2}}$, which determines the location of focal point, is adjustable via reducing $n_2$ by applying negative gate voltage ($n_2$ is the electron density under the top-gated regime; $n_1$ is the density within the hollow area  or raw 2D density).

The functionality of the electronic lens centered between the QPCs can be verified by noticing an enhancement in the detected signal when the focal point aligns with the saddle point of the detector (QPC3). To detect the real focusing signal, the residual signals were \textbf{\textit{simultaneously}} drained to Ohmic contacts 4 and 5 in Fig. 1(b); otherwise, all the injected electrons would be drained via the detector whether or not the focusing condition was matched. Figure 1(c) shows a representative result with QPC 1-3 set to G$_0$ ( G$_0$=$\frac{2e^2}{h}$). The residual current $I_4$+$I_5$ measured at Ohmic contacts 4 and 5 resembled a typical pinched-off behaviour; on the other hand, $I_3$ measured at Ohmic contact 3 yielded a series of peaks. Peaks in $I_3$ near pinched-off regime may arise from charging effect or scattering at low electron density limit as suggested by their insensitivity against transverse magnetic field [Fig.S5 of supplementary information]; the small fluctuation can be a result of universal conductance fluctuation; peak occurred at lens gate voltage $V_{lens}$ $\approx$ -0.90 V [marked by the bold black arrow in Fig. 1(c)] was an indication of electron focusing. The focal length $l_f$ at given $V_{lens}$ can be calculated from lensmaker's equation: $$\frac{1}{l_f}=(N_r - 1) \times [\frac{1}{R_1}-\frac{1}{R_2}+\frac{(N_r-1)D}{N_r R_1 R_2}] \eqno(1)$$ where $R_1$ = 2 $\mu$m and $R_2$ = -2 $\mu$m are the radius of right and left surface of the lens, D = 2 $\mu$m is the thickness of lens, $N_r$ can be extracted from capacitance model\cite{YSU91,YTS18} as $N_r = \sqrt{\frac{V_{pin}}{V_{pin} - V_{lens}}}$, where $V_{pin}$ indicates the pinched-off voltage of the lens gate. Inserting $V_{pin}$ = -1.60 V and $V_{lens}$ = -0.90 V, Eq.1 yielded that $N_r$ = 1.521 and $l_f$ = 2.17 $\mu$m which agrees very well with the lithographically defined distance (2.25 $\mu$m) between geometric centre of the lens and QPC3. The difference in the value may arise from the fact that the effective dimension of the (electrostatic) lens differs from the lithographically defined one; besides, an offset of the lens along primary axis is also possible. An enhancement in $I_3$ happened when the focal point of the electronic lens aligned with the saddle point of QPC3. In addition, the width of the focusing peak also revealed important insights. The focusing peak started forming once the focal point was driven to the vicinity of the entrance of QPC3 and attenuating when the focal point passed the exit; therefore a change in $l_f$ within this range of $V_{lens}$ should match channel length of QPC3. The full-width half maxima (FWHM) of the focusing peak suggested a change in $l_f$ of 313 nm which is consistent with the lithographically defined channel length of 300 nm. The difference in values could probably arise from the finite angular spread of injected electrons.

To further validate the existence of the focusing process, we presented data after the sample was illuminated by a red LED  in Fig. 1(d) with all QPCs set to G$_0$. According to Eq.(1), $N_r$ must remain the same when focusing peak occurred before and after illumination. It was found that the only prominent peak after illumination, the focusing peak, in $I_3$ happened at $V_{len}$ = -1.08 V whereas $V_{pin}$ became -1.90 V as a result of increased 2D electron density, these values suggest $N_r$ = 1.523 just as before illumination. 

It might be concerned that the focusing peak may arise from trivial electrostatic effects such as the cross-coupling between the split gates and lens gate, or coherent effects such as universal conductance fluctuation. These interpretations can be excluded by the data obtained with different injector conductance, different combination of grounding scheme (for instance, both Ohmic 4 and 5 are floating or grounded), temperature and transverse magnetic field dependence [see note 4 of supplementary information for detailed discussion].

\subsection*{Temperature dependence of the focusing peak}   
 
The evolution of focusing peak with lattice temperature elevated from 20 mK to 1.6 K is presented in Fig. 2(a). The focusing peak intensity $\gamma$ weakened with increasing temperature and followed the trend based on electron collimation as shown in Fig. 2(b). To avoid the constant pre-factor which could not be calculated directly, $\gamma (T)$ was normalized against $\gamma (1.6 K)$. $\gamma$ was closely associated with electron collimation, it was suggested that\cite{MMH92} $$\gamma (T)/\gamma (1.6 K) = exp\{-l[l_m (T)^{-1} - l_m (1.6 K)^{-1}]\} \eqno(2)$$  where $l$ is electron propagation length, $l_m(T)$ is the mean free path for electron-electron scattering at given temperature $T$, which is defined as $l_m (T) = v_F \times \tau_{ee} (T)$ where $v_F$ is the Fermi velocity and $\tau_{ee} (T)$ is electron-electron scattering time\cite{MMH92} $$\frac{1}{\tau_{ee} (T)}=\frac{E_F}{h} (\frac{k_B T}{E_F})^2 [ln(\frac{E_F}{k_B T})+ln(\frac{2q_{TF}}{k_F})+1] \eqno(3)$$ where $E_F$ is the Fermi energy, $k_B$ is the Boltzmann constant, $q_{TF}$ is the Thomas-Fermi screening wave vector, $k_F$ is Fermi wave vector. $l$ can be obtained by assuming electrons follow the classical trajectory connecting QPC1 (QPC2) and QPC3 with refraction at the lens taken into account, whereas $l_m(T)$ is fully determined by $T$ and electron density in the area between injectors and detector. The non-uniform electron density underneath the lens (including the top-gated regime and hollow area) was a complex situation. To simplify the calculation, we assume a uniform top-gate, without hollow regime, is patterned between the injector and detectors. The upper bound of the calculation [blue solid line in Fig.2(b)] corresponds to  $V_{lens}$ = -0.90 V applied to the virtual uniform top-gate, whereas lower bound is given by zero-gate voltage [red solid line in Fig.2(b)]. It turned out that the experimental data lied between the lower and upper bound but much closer to the upper bound.   

The  agreement between $l$ and classical trajectory length (through a double-convex lens) further supported that peak at $V_{len}$ = -0.90 V arose from electron focusing. The temperature dependence data also indicate that the focusing peak is not due to trivial coherent effect [supplementary information note 4].

\section*{Focusing peak with asymmetrically gate biased injector }

\begin{figure}
	
	\includegraphics[height=2.0in,width=3.4in]{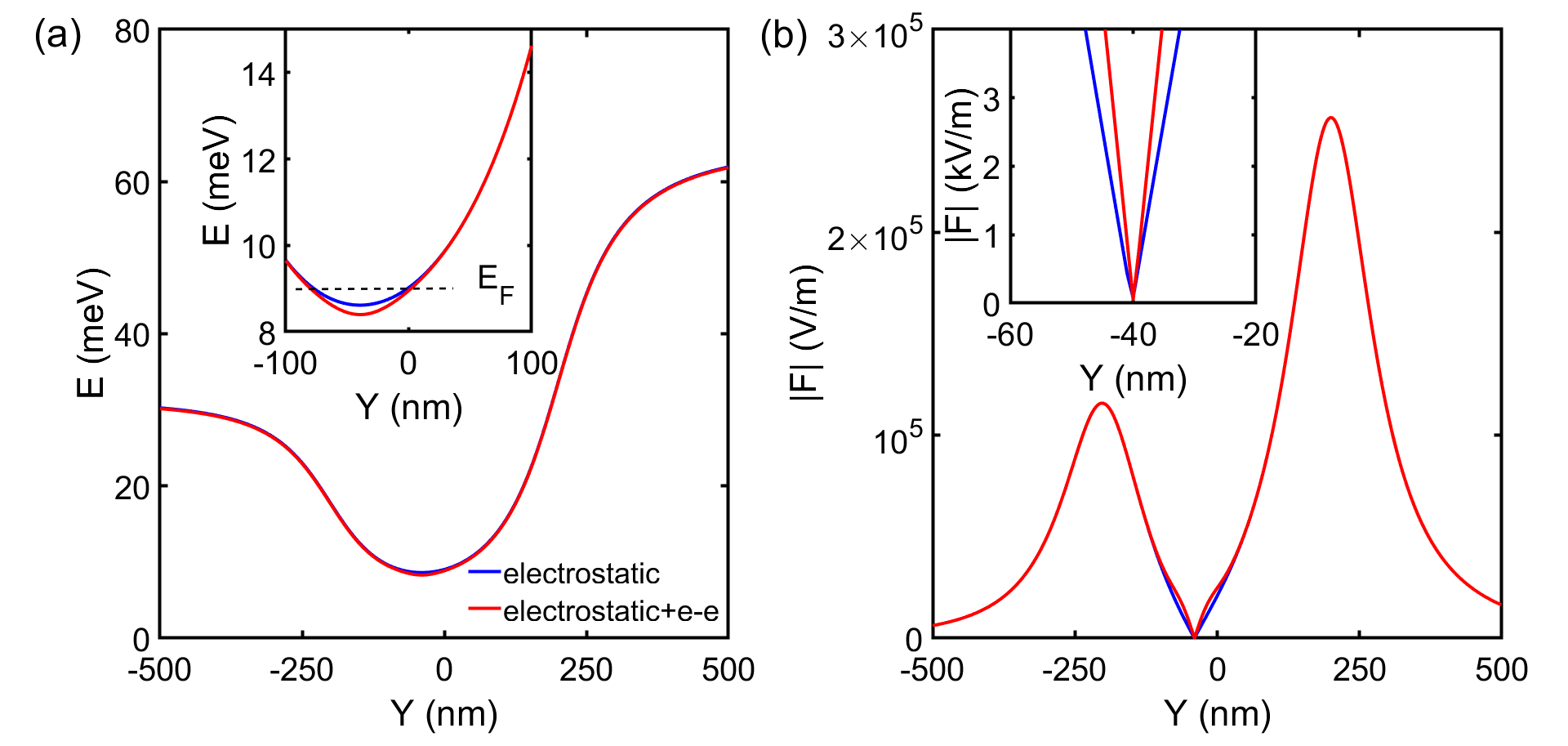}
	
	\caption{Simulated confinement potential and electric field. \textbf{(a)} and \textbf{(b)} show simulated confinement potential and absolute value of electric field with $\Delta V$ = -1 V. The blue trace considers electrostatic contribution only whereas the red trace is with correction due to electron-electron interaction. Inset highlights results around the potential minimum.}  
	
	\label{fig:4}
	
\end{figure}  
      
After confirming the lens was capable of focusing electrons, we utilized the technique to investigate the property of an asymmetrically gate biased QPC where lateral spin-orbit coupling (LSOC) was suggested to play a vital role\cite{DRW09,BKF12, CHS15,VSO16,DJC17}, even in GaAs\cite{PPZ18}.

The behaviour of injector QPCs with asymmetric gate bias was characterized using standard conductance measurement. Taking QPC1 as an example, if gate voltage $V_{sg}$ was applied to the bottom arm of the split gate [refers to Fig.1 (b)], $V_{sg}+\Delta V$ would be applied to the upper arm. $\Delta V$ was kept negative so that the 1D channel shifted towards the primary axis [Fig 1(a)] instead of the tips of the lens, to avoid diffraction at the tips\cite{HHH00}. A short plateau-like feature, as marked by the red arrow in Fig.3(a), below 0.7-anomaly started forming with increasing $\Delta V$ similar to previous observations\cite{DRW09,BKF12,CHS15,VSO16,DJC17,PPZ18}. We denoted this feature as sub-0.7-anomaly hereafter. However, the sub-0.7-anomaly was unlikely to arise from previously proposed LSOC\cite{DRW09,BKF12,CHS15,VSO16,DJC17,PPZ18} . In GaAs electron gas, the ground-state electrons propagate along the potential minimum of the 1D channel. For the studied device, the lateral electric field at the potential minimum is about 100 V/m with $\Delta V$ = -1 V according to simulation in Fig. 4 [supplementary information note 8], which is insufficient to generate noticeable LSOC. Ideally speaking the electric field at the potential minimum should be zero, the finite value obtained here is likely due to the finite grid spacing used in the simulation.

Additional insight into the observed sub-0.7-anomaly can be extracted from the focusing experiment. In this experiment QPC1 (QPC2 was not used) was set to  G$_0$ under different $\Delta V$; on the other hand, QPC3 was under symmetric bias and fixed at G$_0$. The results are summarized in Fig. 3(b) and (c). It was clear that the focusing peak centred at -1.08 V regardless of $\Delta V$. The robustness of focusing peak position indicates that electron injection angle from a QPC tends to concentrate at 0$^\circ$ even with asymmetric confinement. Assuming that the injection angle concentrates at a non-zero angle $\alpha$, then the injected electrons should be guided to $f^\ast$ instead of the focal point [see upper panel of Fig. 1(a)]; to make $f^\ast$ align with the saddle point of detector, an adjustment on $V_{lens}$ is necessary, for instance the focusing peak should occur at $V_{lens}$ = -1.17 V assuming $\alpha$ = 6$^\circ$. Although the central position of focusing peak showed no explicit dependence on $\Delta V$, FWHM of focusing peak almost increased monotonically against $\Delta V$, as shown in Fig. 3(c). The broadening of FWHM suggests a larger angular spread of injected electrons; in other words, a reduction in collimation [supplementary information note 7]. FWHM started increasing rapidly at $\Delta V$ = -0.4 V which was roughly the same  $\Delta V$ to trigger the sub-0.7-anomaly.

We suggest that the occurrence of sub-0.7-anomaly and increase in FWHM of focusing peak are possibly due to electron-electron interaction. Applying the asymmetric gate-bias $\Delta V$ results in a change in the effective length of the 1D channel, whereas the electron-electron interaction (e-e interaction) especially the exchange part is sensitive to the channel length, it has been shown conductance anomaly can occur between 0.8$\times$ $\frac{2e^2}{h}$ to 0.4$\times$ $\frac{2e^2}{h}$ depends on the channel length\cite{JYB06}. On the other hand, for a symmetrically gate-biased QPC, the angle $\alpha$ within which the electrons are highly collimated is given by\cite{BH89,RCC00},$$ \alpha = \pm arcsin(\sqrt{\frac{E_F - E_b}{E_F}} \times \frac{W_{min}}{W_{max}}) \eqno(4)$$ where $E_b$ is the potential at the saddle point where minimum 1D channel width $W_{min}$ occurs, $W_{max}$ is the critical channel width where electron transport still remains non-adiabatic. It is found that $\alpha$ changes from 5.24$^\circ$ ($\Delta$V = 0) to 5.17$^\circ$ ($\Delta$V = -1 V) without e-e interaction; on the other hand, $\alpha$ increases from 5.77$^\circ$ ($\Delta$V = 0) to 6.77$^\circ$ ($\Delta$V = -1 V) after taking e-e interaction into account. It seems that e-e interaction is an essential ingredient for an observable change in FWHM [see note 9 of supplementary information for a more detailed discussion].

It is also important to check the role of disorder, which can result in multiple irregular features on conductance characteristic. If we try to understand the conductance measurement in Fig.3 (a) based on disorder, it is natural to think the magenta trace (the third one from the right side), where several irregular features are observed, corresponds to the case where the effect of disorder is most substantial. The smoothing of the two most right traces indicates the channel is moving away from the disorder. It is then expected that the FWHM of focusing peak should follow a non-monotonic trend if disorder plays a primary role, however, the experimental result shows a monotonic trend.

\section*{Extending the focusing scheme to material with strong spin-orbit coupling} 

It is helpful to apply the focusing scheme to material with  strong intrinsic spin-orbit coupling (SOC). The strong intrinsic SOC can causes lateral motion, so that the two spin-branches tend to move along the opposite edges of the 1D channel where the lateral electric field is noticeable\cite{DRW09}. LSOC is induced as a result of the lateral electric field.      

It has been shown that a small out-plane magnetic field can make angular distribution centered at a finite angle rather than 0$^\circ$\cite{UTK94} . The effective out-plane magnetic field induced by intrinsic SOC can have similar influence. Therefore, it is expected that the angular spread has two peaks at $\pm \theta$ in the presence of intrinsic SOC ($\pm$ sign depends on spin orientation) instead of a single peak at 0$^\circ$ when intrinsic SOC is absent. The spin branches will thus lead to two focusing peaks. Information on both the intrinsic SOC and LSOC can be extracted from the focusing profile.

\section*{Conclusion} 

We have developed a double-convex electron focusing lens which is an essential component for a complete tool kit of electron optics. A focusing peak occurred when the focal point of the lens aligned with the detector, the intensity of focusing was closely associated with the degree of electron collimation. Using the focusing lens, we found that the injection angle of 1D electrons tends to concentrate at 0$^\circ$ even with considerable asymmetric gate bias $\Delta V$. However, the angular spread broadened with increasing $\Delta V$. The increment of FWHM was correlated with the occurrence of a sub-0.7-anomaly, possibly due to electron-electron interaction. The focusing scheme is ready to be extended to material with strong intrinsic spin-orbit interaction, where it allows to selectively polarized electron spin in the injector/detector and study the interplay between local spin states by monitoring the evolution in corresponding focusing peak.

\textbf{Acknowledgments}\\

The authors gratefully acknowledge assistance on low-temperature measurement setup from Sanjeev Kumar. This work was supported by the Engineering and Physical Sciences Research Council (EPSRC), U.K.

\textbf{Author contributions}\\

CY designed and performed the experiments; CY and MP analyzed the data; PS fabricated the devices; JG did e-beam lithography; IF and DR grew the wafer; CY wrote the paper with inputs from MP, and other co-authors. All authors read and approved the final manuscript.

\textbf{Additional Information}\\

Competing Interests: The authors declare no competing interests.


\begin{thebibliography}{10}
	
\section*{References} 

	\expandafter\ifx\csname url\endcsname\relax
	\def\url#1{\texttt{#1}}\fi
	\expandafter\ifx\csname urlprefix\endcsname\relax\def\urlprefix{URL }\fi
	\providecommand{\bibinfo}[2]{#2}
	\providecommand{\eprint}[2][]{\url{#2}}
	
	\bibitem{CFX05}
	\bibinfo{author}{Crooker, S.} \emph{et~al.}
	\newblock \bibinfo{title}{Imaging spin transport in lateral
		ferromagnet/semiconductor structures}.
	\newblock \emph{\bibinfo{journal}{Science}} \textbf{\bibinfo{volume}{309}},
	\bibinfo{pages}{2191--2195} (\bibinfo{year}{2005}).
	
	\bibitem{WPI10}
	\bibinfo{author}{Wunderlich, J.} \emph{et~al.}
	\newblock \bibinfo{title}{Spin Hall effect transistor}.
	\newblock \emph{\bibinfo{journal}{Science}} \textbf{\bibinfo{volume}{330}},
	\bibinfo{pages}{1801--1804} (\bibinfo{year}{2010}).
	
	\bibitem{BDS12}
	\bibinfo{author}{Betthausen, C.} \emph{et~al.}
	\newblock \bibinfo{title}{Spin-transistor action via tunable Landau-Zener
		transitions}.
	\newblock \emph{\bibinfo{journal}{Science}} \textbf{\bibinfo{volume}{337}},
	\bibinfo{pages}{324--327} (\bibinfo{year}{2012}).
	
	\bibitem{RTB07}
	\bibinfo{author}{Rycerz, A.}, \bibinfo{author}{Tworzyd{\l}o, J.} \&
	\bibinfo{author}{Beenakker, C.}
	\newblock \bibinfo{title}{Valley filter and valley valve in graphene}.
	\newblock \emph{\bibinfo{journal}{Nature Physics}}
	\textbf{\bibinfo{volume}{3}}, \bibinfo{pages}{172} (\bibinfo{year}{2007}).
	
	\bibitem{SYC16}
	\bibinfo{author}{Schaibley, J.~R.} \emph{et~al.}
	\newblock \bibinfo{title}{Valleytronics in 2D materials}.
	\newblock \emph{\bibinfo{journal}{Nature Reviews Materials}}
	\textbf{\bibinfo{volume}{1}}, \bibinfo{pages}{16055} (\bibinfo{year}{2016}).
	
	\bibitem{HDD04}
	\bibinfo{author}{Hill, M.~T.} \emph{et~al.}
	\newblock \bibinfo{title}{A fast low-power optical memory based on coupled
		micro-ring lasers}.
	\newblock \emph{\bibinfo{journal}{nature}} \textbf{\bibinfo{volume}{432}},
	\bibinfo{pages}{206} (\bibinfo{year}{2004}).
	
	\bibitem{PAL04}
	\bibinfo{author}{Babichev, S.~A.}, \bibinfo{author}{Appel, J.} \&
	\bibinfo{author}{Lvovsky, A.~I.}
	\newblock \bibinfo{title}{Homodyne tomography characterization and nonlocality
		of a dual-mode optical qubit}.
	\newblock \emph{\bibinfo{journal}{Phys. Rev. Lett.}}
	\textbf{\bibinfo{volume}{92}}, \bibinfo{pages}{193601}
	(\bibinfo{year}{2004}).
	
	\bibitem{TLS00}
	\bibinfo{author}{Topinka, M.} \emph{et~al.}
	\newblock \bibinfo{title}{Imaging coherent electron flow from a quantum point
		contact}.
	\newblock \emph{\bibinfo{journal}{Science}} \textbf{\bibinfo{volume}{289}},
	\bibinfo{pages}{2323--2326} (\bibinfo{year}{2000}).
	
	\bibitem{DD90}
	\bibinfo{author}{Datta, S.} \& \bibinfo{author}{Das, B.}
	\newblock \bibinfo{title}{Electronic analog of the electro‐optic modulator}.
	\newblock \emph{\bibinfo{journal}{Applied Physics Letters}}
	\textbf{\bibinfo{volume}{56}}, \bibinfo{pages}{665--667}
	(\bibinfo{year}{1990}).
	
	\bibitem{DTW01}
	\bibinfo{author}{Duncan, D.~S.}, \bibinfo{author}{Topinka, M.~A.},
	\bibinfo{author}{Westervelt, R.~M.}, \bibinfo{author}{Maranowski, K.~D.} \&
	\bibinfo{author}{Gossard, A.~C.}
	\newblock \bibinfo{title}{Aharonov-Bohm phase shift in an open electron
		resonator}.
	\newblock \emph{\bibinfo{journal}{Phys. Rev. B}} \textbf{\bibinfo{volume}{64}},
	\bibinfo{pages}{033310} (\bibinfo{year}{2001}).
	
	\bibitem{SHU90}
	\bibinfo{author}{Sivan, U.}, \bibinfo{author}{Heiblum, M.},
	\bibinfo{author}{Umbach, C.~P.} \& \bibinfo{author}{Shtrikman, H.}
	\newblock \bibinfo{title}{Electrostatic electron lens in the ballistic regime}.
	\newblock \emph{\bibinfo{journal}{Phys. Rev. B}} \textbf{\bibinfo{volume}{41}},
	\bibinfo{pages}{7937--7940} (\bibinfo{year}{1990}).
	
	\bibitem{SSB90}
	\bibinfo{author}{Spector, J.}, \bibinfo{author}{Stormer, H.},
	\bibinfo{author}{Baldwin, K.}, \bibinfo{author}{Pfeiffer, L.} \&
	\bibinfo{author}{West, K.}
	\newblock \bibinfo{title}{Electron focusing in two-dimensional systems by means
		of an electrostatic lens}.
	\newblock \emph{\bibinfo{journal}{Applied physics letters}}
	\textbf{\bibinfo{volume}{56}}, \bibinfo{pages}{1290--1292}
	(\bibinfo{year}{1990}).
	
	\bibitem{MSB90}
	\bibinfo{author}{Molenkamp, L.~W.} \emph{et~al.}
	\newblock \bibinfo{title}{Electron-beam collimation with a quantum point
		contact}.
	\newblock \emph{\bibinfo{journal}{Phys. Rev. B}} \textbf{\bibinfo{volume}{41}},
	\bibinfo{pages}{1274--1277} (\bibinfo{year}{1990}).
	
	\bibitem{STO92}
	\bibinfo{author}{Saito, M.}, \bibinfo{author}{Takatsu, M.},
	\bibinfo{author}{Okada, M.} \& \bibinfo{author}{Yokoyama, N.}
	\newblock \bibinfo{title}{Analysis of the angular distribution of electrons
		injected through a quantum point contact by use of a Green's function with a
		weak-magnetic-field approximation}.
	\newblock \emph{\bibinfo{journal}{Phys. Rev. B}} \textbf{\bibinfo{volume}{46}},
	\bibinfo{pages}{13220--13233} (\bibinfo{year}{1992}).
	
	\bibitem{SI19}
	\bibinfo{title}{The supplementary information includes results under different
		experimental conditions and details on simulation, also ref. 30-33} .
	
	\bibitem{YSU91}
	\bibinfo{author}{Yacoby, A.}, \bibinfo{author}{Sivan, U.},
	\bibinfo{author}{Umbach, C.~P.} \& \bibinfo{author}{Hong, J.~M.}
	\newblock \bibinfo{title}{Interference and dephasing by electron-electron
		interaction on length scales shorter than the elastic mean free path}.
	\newblock \emph{\bibinfo{journal}{Phys. Rev. Lett.}}
	\textbf{\bibinfo{volume}{66}}, \bibinfo{pages}{1938--1941}
	(\bibinfo{year}{1991}).
	
	\bibitem{YTS18}
	\bibinfo{author}{Yan, C.} \emph{et~al.}
	\newblock \bibinfo{title}{Coherent spin amplification using a beam splitter}.
	\newblock \emph{\bibinfo{journal}{Phys. Rev. Lett.}}
	\textbf{\bibinfo{volume}{120}}, \bibinfo{pages}{137701}
	(\bibinfo{year}{2018}).
	
	\bibitem{MMH92}
	\bibinfo{author}{Molenkamp, L.~W.}, \bibinfo{author}{Brugmans, M. J.~P.},
	\bibinfo{author}{van Houten, H.} \& \bibinfo{author}{Foxon, C.~T.}
	\newblock \bibinfo{title}{Electron-electron scattering probed by a collimated
		electron beam}.
	\newblock \emph{\bibinfo{journal}{Semiconductor Science and Technology}}
	\textbf{\bibinfo{volume}{7}}, \bibinfo{pages}{B228} (\bibinfo{year}{1992}).
	
	\bibitem{DRW09}
	\bibinfo{author}{Debray, P.} \emph{et~al.}
	\newblock \bibinfo{title}{All-electric quantum point contact spin-polarizer}.
	\newblock \emph{\bibinfo{journal}{Nature Nanotechnology}}
	\textbf{\bibinfo{volume}{4}}, \bibinfo{pages}{759} (\bibinfo{year}{2009}).
	
	\bibitem{BKF12}
	\bibinfo{author}{Burke, A.} \emph{et~al.}
	\newblock \bibinfo{title}{Extreme sensitivity of the spin-splitting and 0.7
		anomaly to confining potential in one-dimensional nanoelectronic devices}.
	\newblock \emph{\bibinfo{journal}{Nano letters}} \textbf{\bibinfo{volume}{12}},
	\bibinfo{pages}{4495--4502} (\bibinfo{year}{2012}).
	
	\bibitem{CHS15}
	\bibinfo{author}{Chuang, P.} \emph{et~al.}
	\newblock \bibinfo{title}{All-electric all-semiconductor spin field-effect
		transistors}.
	\newblock \emph{\bibinfo{journal}{Nature nanotechnology}}
	\textbf{\bibinfo{volume}{10}}, \bibinfo{pages}{35} (\bibinfo{year}{2015}).
	
	\bibitem{VSO16}
	\bibinfo{author}{von Pock, J.} \emph{et~al.}
	\newblock \bibinfo{title}{Quantization and anomalous structures in the
		conductance of Si/SiGe quantum point contacts}.
	\newblock \emph{\bibinfo{journal}{Journal of Applied Physics}}
	\textbf{\bibinfo{volume}{119}}, \bibinfo{pages}{134306}
	(\bibinfo{year}{2016}).
	
	\bibitem{DJC17}
	\bibinfo{author}{Das, P.~P.} \emph{et~al.}
	\newblock \bibinfo{title}{Dependence of the 0.5 X 2e$^2$/h  conductance plateau on the aspect ratio of InAs quantum point contacts with in-plane side gates}.
	\newblock \emph{\bibinfo{journal}{Journal of Applied Physics}}
	\textbf{\bibinfo{volume}{121}}, \bibinfo{pages}{083901}
	(\bibinfo{year}{2017}).
	
	\bibitem{PPZ18}
	\bibinfo{author}{Pokhabov, D.} \emph{et~al.}
	\newblock \bibinfo{title}{Lateral-electric-field-induced spin polarization in a
		suspended GaAs quantum point contact}.
	\newblock \emph{\bibinfo{journal}{Applied Physics Letters}}
	\textbf{\bibinfo{volume}{112}}, \bibinfo{pages}{082102}
	(\bibinfo{year}{2018}).
	
	\bibitem{HHH00}
	\bibinfo{author}{Hersch, J.~S.}, \bibinfo{author}{Haggerty, M.~R.} \&
	\bibinfo{author}{Heller, E.~J.}
	\newblock \bibinfo{title}{Influence of diffraction on the spectrum and wave
		functions of an open system}.
	\newblock \emph{\bibinfo{journal}{Phys. Rev. E}} \textbf{\bibinfo{volume}{62}},
	\bibinfo{pages}{4873--4888} (\bibinfo{year}{2000}).
	
	\bibitem{JYB06}
	\bibinfo{author}{Jaksch, P.}, \bibinfo{author}{Yakimenko, I.} \&
	\bibinfo{author}{Berggren, K.-F.}
	\newblock \bibinfo{title}{From quantum point contacts to quantum wires:
		Density-functional calculations with exchange and correlation effects}.
	\newblock \emph{\bibinfo{journal}{Phys. Rev. B}} \textbf{\bibinfo{volume}{74}},
	\bibinfo{pages}{235320} (\bibinfo{year}{2006}).
	
	\bibitem{BH89}
	\bibinfo{author}{Beenakker, C. W.~J.} \& \bibinfo{author}{Houten, H.~v.}
	\newblock \bibinfo{title}{Magnetotransport and nonadditivity of point-contact
		resistances in series}.
	\newblock \emph{\bibinfo{journal}{Phys. Rev. B}} \textbf{\bibinfo{volume}{39}},
	\bibinfo{pages}{10445--10448} (\bibinfo{year}{1989}).
	
	\bibitem{RCC00}
	\bibinfo{author}{Crook, R.}, \bibinfo{author}{Smith, C.~G.},
	\bibinfo{author}{Barnes, C. H.~W.}, \bibinfo{author}{Simmons, M.~Y.} \&
	\bibinfo{author}{Ritchie, D.~A.}
	\newblock \bibinfo{title}{Imaging diffraction-limited electronic collimation
		from a non-equilibrium one-dimensional ballistic constriction}.
	\newblock \emph{\bibinfo{journal}{Journal of Physics: Condensed Matter}}
	\textbf{\bibinfo{volume}{12}}, \bibinfo{pages}{L167--L172}
	(\bibinfo{year}{2000}).
	
	\bibitem{UTK94}
	\bibinfo{author}{Usuki, T.}, \bibinfo{author}{Takatsu, M.},
	\bibinfo{author}{Kiehl, R.~A.} \& \bibinfo{author}{Yokoyama, N.}
	\newblock \bibinfo{title}{Numerical analysis of electron-wave detection by a
		wedge-shaped point contact}.
	\newblock \emph{\bibinfo{journal}{Phys. Rev. B}} \textbf{\bibinfo{volume}{50}},
	\bibinfo{pages}{7615--7625} (\bibinfo{year}{1994}).
	
	\bibitem{OBH10}
	\bibinfo{author}{Ofek, N.} \emph{et~al.}
	\newblock \bibinfo{title}{Role of interactions in an electronic
		fabry{\textendash}perot interferometer operating in the quantum hall effect
		regime}.
	\newblock \emph{\bibinfo{journal}{Proceedings of the National Academy of
			Sciences}}  (\bibinfo{year}{2010}).
	
	\bibitem{DLS95}
	\bibinfo{author}{Davies, J.~H.}, \bibinfo{author}{Larkin, I.~A.} \&
	\bibinfo{author}{Sukhorukov, E.~V.}
	\newblock \bibinfo{title}{Modelling the patterned two‐dimensional electron
		gas: Electrostatics}.
	\newblock \emph{\bibinfo{journal}{Journal of Applied Physics}}
	\textbf{\bibinfo{volume}{77}}, \bibinfo{pages}{4504--4512}
	(\bibinfo{year}{1995}).
	
	\bibitem{LJN92}
	\bibinfo{author}{Martin-Moreno, L.}, \bibinfo{author}{Nicholls, J.~T.},
	\bibinfo{author}{Patel, N.~K.} \& \bibinfo{author}{Pepper, M.}
	\newblock \bibinfo{title}{Non-linear conductance of a saddle-point
		constriction}.
	\newblock \emph{\bibinfo{journal}{Journal of Physics: Condensed Matter}}
	\textbf{\bibinfo{volume}{4}}, \bibinfo{pages}{1323} (\bibinfo{year}{1992}).
	
	\bibitem{WB96}
	\bibinfo{author}{Wang, C.-K.} \& \bibinfo{author}{Berggren, K.-F.}
	\newblock \bibinfo{title}{Spin splitting of subbands in quasi-one-dimensional
		electron quantum channels}.
	\newblock \emph{\bibinfo{journal}{Phys. Rev. B}} \textbf{\bibinfo{volume}{54}},
	\bibinfo{pages}{R14257--R14260} (\bibinfo{year}{1996}).
	
\end{thebibliography}
\end{document}